\documentclass{article}
\usepackage[utf8]{inputenc}
\usepackage[margin=1in]{geometry}
\usepackage{todonotes}
\usepackage{arydshln}
\usepackage{natbib}
\usepackage{footnotebackref}
\usepackage{authblk}
\usepackage{setspace}

\providecommand{\keywords}[1]
{
  \small	
  \textbf{\textit{Keywords---}} #1
}

\title{
\textbf{System Safety and Artificial Intelligence\footnote{To appear in the Oxford Handbook on AI Governance (2022)}}}
\author[1]{Roel I.J. Dobbe}
\affil[1]{Faculty of Technology, Policy and Management, Delft University of Technology, The Netherlands, \url{r.i.j.dobbe@tudelft.nl}.}
\date{January 2022}

\begin{document}

\maketitle

\begin{abstract}
\onehalfspacing
This chapter formulates seven lessons for preventing harm in artificial intelligence (AI) systems based on insights
from the field of system safety for software-based automation in safety-critical domains. 
New applications of AI across societal domains and public organizations and infrastructures come with new hazards, which lead to new forms of harm, both grave and pernicious.
The text addresses the lack of consensus for diagnosing and eliminating new AI system hazards.
For decades, the field of \emph{system safety} has dealt with accidents and harm in safety-critical systems governed by varying degrees of software-based automation and decision-making.
This field embraces the core assumption of \emph{systems and control} that AI systems cannot be safeguarded by technical design choices on the model or algorithm alone, instead requiring an end-to-end hazard analysis and design frame that includes the context of use, impacted stakeholders and the formal and informal institutional environment in which the system operates.
Safety and other values are then inherently \emph{socio-technical and emergent system properties} that require design and control measures to instantiate these across the technical, social and institutional components of a system.
This chapter honors system safety pioneer Nancy Leveson, by situating her core lessons for today's AI system safety challenges.
For every lesson, concrete tools are offered for rethinking and reorganizing the safety management of AI systems, both in design and governance.
This history tells us that effective AI safety management requires transdisciplinary approaches and a shared language that allows involvement of all levels of society.
\end{abstract}





\section{Introduction}


The emergence of AI systems in public services as well as in public spaces and infrastructures, has led to a plethora of new hazards, leading to accidents with fatal consequences~\citep{raji_concrete_2020} and increasing concerns over the risks posed to democratic institutions~\citep{crawford_ai_2019}.
The criteria for identifying and diagnosing safety risks in complex social contexts remain unclear and contested. While various proposals have emerged specifying \emph{what} it is we need to strive for in terms of dealing with ethical, legal and societal implications of AI systems, there still is a long way to go to understand \emph{how} to translate such higher-level principles and requirements to the development as well as the operation and governance of these systems in practice~\citep{mittelstadt_principles_2019,dobbe_hard_2021}.
A system perspective is needed~\citep{zuiderwijk_implications_2021} that facilitates the identification of (new) complex safety hazards in ways that do justice to those bearing the brunt of AI systems~\citep{benjamin_race_2020},
and can contribute to strengthening the rule of law.

While there is still considerable disagreement on what entails an appropriate definition of AI, recent policy efforts led by the OECD have converged on a seeing AI as a \emph{sociotechnical system with a complex lifecycle}~\citep{oecd_recommendation_2021}.
Here we define sociotechnical AI systems as consisting of technical AI artefacts, human agents, and institutions, in which the AI artefact influences its real or virtual environment by automating, supporting or augmenting decision-making.
The technical AI components in such a system may vary from logic or knowledge-based models to machine learning models to statistical or Bayesian to search or optimization methods. 
In systems engineering, safety is understood as an \emph{emergent} property, which can only be instantiated and controlled for across the above-mentioned system elements~\citep{leveson_engineering_2012}.


Core to the establishment of a system perspective are lessons from decades of knowledge built up about what constitutes safety in systems subject to software-based automation.
These lessons have become central to the organization of many domains and markets, such as in aviation and healthcare.
However, scanning both AI systems literature as well as policy proposals, it is evident that these lessons have yet to be absorbed.
In this chapter, I provide an on-ramp into the system safety canon for different disciplines and people involved or invested in the safeguarding of AI systems.
I center the seminal work of Nancy Leveson, a pioneer of system safety in engineering systems. 
In her magnum opus \emph{Engineering a safer world: systems thinking applied to safety}, Leveson draws seven lessons based on recurring issues that hinder the safeguarding of complex systems subject to forms of soft-ware based automation~\citep{leveson_engineering_2012}.
I coin these the \emph{Leveson Lessons} and interpret them for a variety of concrete challenges that are emerging for safeguarding AI systems in different domains.

\section{Leveson's Lessons for AI System Safety}

In this section, the seven Leveson lessons each inform a core implication for the emerging field of research and practice in AI system development and governance, and some concrete examples of approaches and tools from system safety that can form on-ramps and inspiration for \emph{how} to put principles for safe AI systems into practice. 
Table~1 provides an overview of all lessons, implications and example system safety strategies.
These tools should not be taken as a comprehensive fix for AI system safety challenges, but as a starting point to overcome gaps in mindset across policy makers, system designers and managers, and to ensure stakeholders implicated by potential hazards have a voice in efforts to safeguard AI systems.

\subsection{Shift Focus from Component Reliability to System Hazard Elimination}
\label{sec:systemhazards}
\noindent \textit{\textbf{Leveson Lesson 1:} High reliability is neither necessary nor sufficient for safety.
}

Recent work in the area of ``AI Safety'' has largely focused its efforts on the internal technical components and assumptions of the AI subsystem~\citep{amodei_concrete_2016}.
This formulation of AI systems puts most of its emphasis on the mathematical formulation, including the objective or reward function, the model class with its input variables and data, its parameters and the task it is trying to model. 
As a result, it does not comprehensively address how a system is used in practice and interacts with other (human) agents, systems and its broader environment.
%
In a recent paper, Raji and Dobbe show that a focus on technical components alone cannot explain AI system accidents~\citep{raji_concrete_2020}.
They consider the use of machine learning models to predicts potential criminal activity. To ensure the resulting \emph{predictive policing system} is ``fairness'', various technical efforts propose metrics and tools to prevent some form of bias in the model's outputs. 
However, when the model is used by inherently biased police practices~\citep{richardson_dirty_2019}, the discretion of police forces will still lead to more arrests in certain over-policed neighbourhoods~\citep{lum_predict_2016}. 
Data collected from such practices can then be used to retrain the model or test new systems, leading to forms of \emph{emergent} bias and discrimination that can't be prevented by the logical components of the AI subsystem~\citep{dobbe_broader_2018}. 

\subsubsection*{Relevant System Safety Strategy: Identify hazards at the systems rather than component level}

Rather than being able to be decomposed into components and then examined and designed for safety, AI systems are situated in complex contexts and interact with other components, both in terms of human agents, social organizations and other technical systems. 
System theorists call the behavior that such systems exhibit \emph{organized complexity}~\citep{leveson_engineering_2012}.
Typically, these systems are too complex for the sole use of ``divide and conquer'' analysis and design common in the traditional (engineering) sciences. In addition, such systems are not regular and random enough in their behavior to be studied purely statistically (which we return to in Section~\ref{sec:mechanisms}).

In Leveson's system safety perspective, a \emph{hazard} is carefully defined as ``[a] system state or set of conditions that, together with a particular set of worst-case environmental conditions, will lead to an accident (loss).''~\citep[p.184]{leveson_engineering_2012}
This definition limits the focus to states the system should never be in and gives designers greater freedom and ability to design hazards out of the system.  

These fundamentals provide two important insights. Firstly, it makes most sense to draw system boundaries in ways that include conditions related to accidents over which system designers have some control.
This enables the system designer to translate identified hazards into concrete requirements, typically formulated as a constraint on design or operation of the system. 
We will address these in Section~\ref{sec:accidentmodel}. 

\begin{figure}[h]
    \center{\includegraphics[width=10cm]
    {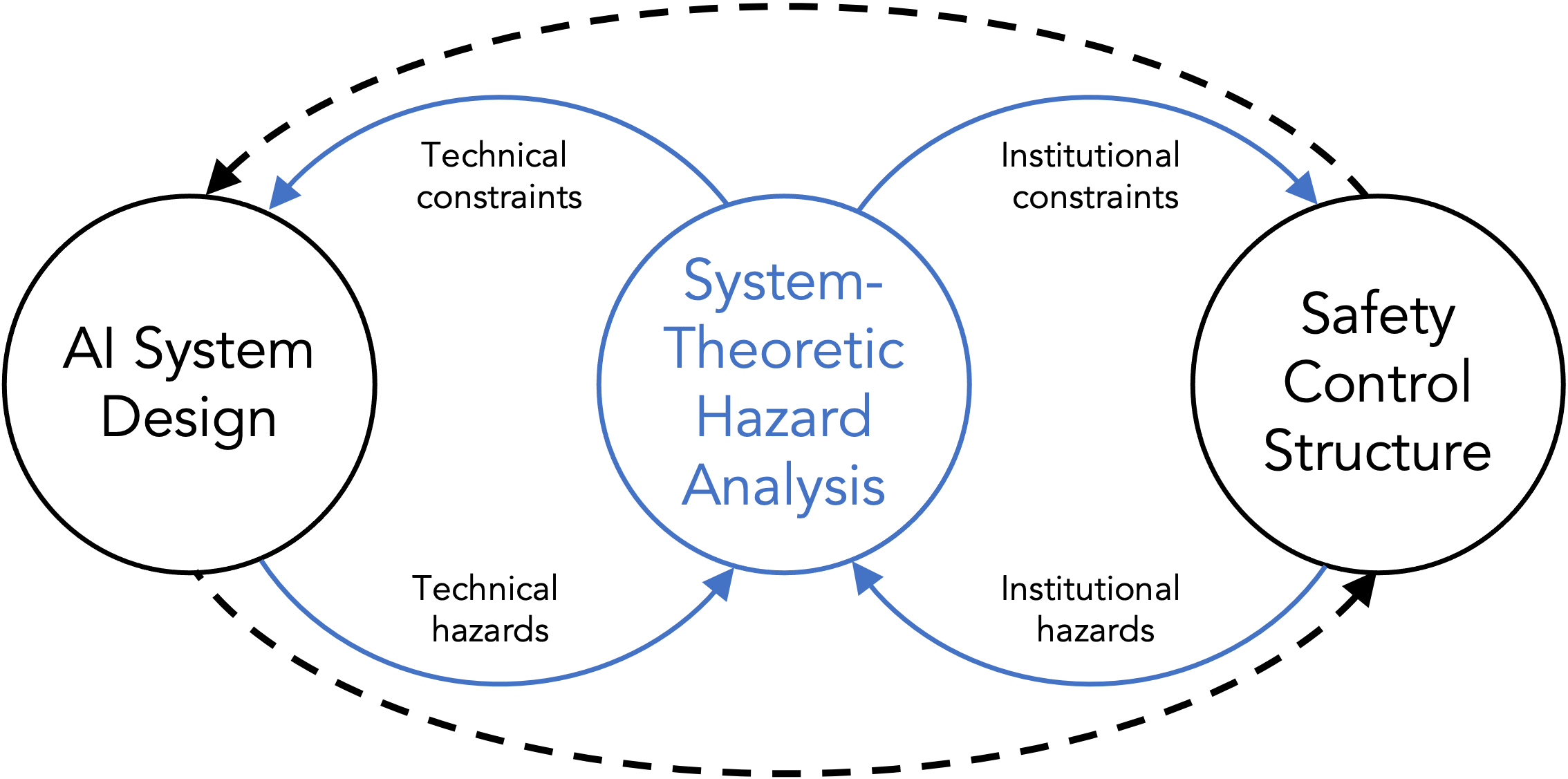}}
    \caption{\label{fig:safetyguideddesign} Hazard analysis as integrally informing the design of AI systems (in black), as well as the design of the institutional safety control structure (in blue). Adapted from~\citep[Chapter 9]{leveson_engineering_2012}.}
\end{figure}

Secondly, system developers typically will not have control over all conditions related to accidents.
Therefore, in order for the identification and elimination of hazards to be complete and properly assigned in terms of responsibilities, one needs to also address the institutional context.
In system safety, institutional design is referred to as the \emph{safety control structure}, which will be further discussed in Section~\ref{sec:operations}.
The need to integrate hazard analysis in both system design as well as operation and the broader institutional safety control structure is depicted in Figure~\ref{fig:safetyguideddesign}.

Let's return to the predictive policing system example.
In order to properly identify the discriminatory nature of a predictive policing system, the system designer may draw the system boundary to include the contexts from which the data originates and in which the AI-tool is used.
This would naturally include the issue of institutional discrimination related to the use of these tools in the system-theoretic safety analysis and prevent naive application of these tools in practice.

\subsection{Shift from Event-based to Constraint-based Accident Models}
\label{sec:accidentmodel}
\noindent \textit{\textbf{Leveson Lesson 2:} Accidents are complex processes involving the entire sociotechnical system. Traditional event-chain models cannot describe this process adequately.
}

Accidents are often studied and attempted to be explained in terms of ``root causes''. In doing so, many accident models rely on identifying chains of causal events. 
Leveson argues that a focus on causal event chains tends to narrowly isolate technical factors, engineering activities and operator errors, thereby overlooking systemic factors that could inform prevention of future accidents. 

A similar critique has surfaced recently for causal and counterfactual methods that aim to capture the ``data generating process'' of machine learning models, in an effort to align model-based decisions with causal structures. 
~\citet{hu_whats_2020} show how such approaches tend to mistake \emph{constitutive} relationships for causal relationships in an effort to answer \emph{why} questions. 
Constitutive relationships comprise variables that together constitute a category, but that do not have a clear temporal relationship, and therefore cannot be modeled as causal event-chains.
\citet{barocas_hidden_2020} provide a more comprehensive account of the limitations of counterfactual methods proposed for improving the quality and fairness of decisions supported or automated by machine learning models, showing that counterfactual insights do not necessarily map to viable actions in the real-world and overlook other dimensions of the decision subject or process.

\subsubsection*{Relevant System Safety Strategy: Ensure safety through socio-technical constraints}

Rather than trying to capture and explain safety concerns with causality,  
a systems-theoretic view uses \emph{constraints} on a socio-technical system's components and their interactions, and studies and designs how and to what extent these can be effectively controlled for.
Here, \emph{control} should be interpreted to not only the role of operators or automation, but can also be attained by integrating constraints in the (physical) design of the system or through social forms of control, which may include organizational, governmental, and regulatory structures (such as rules for how a system can or cannot be used), but they may also be cultural (such as building a safe culture to report issues that could lead to safety hazards)~\citep{leveson_engineering_2012}.



Here, we look at an autonomous vehicle example to see how system-level constraints can be identified, and how responsibility for enforcing these must be divided up and allocated to appropriate stakeholders.
On May 7, 2016, a Tesla Model S was using  using its ``Autopilot'' feature and crashed near Williston, Florida, leading to the death of the driver Joshua Brown. 
The car did not break when a tractor trailer drove across the highway perpendicularly. 
In its initial news release, Tesla stated that ``[n]either autopilot nor the driver noticed the white side of the tractor-trailer against a brightly lit sky, so the brake was not applied,'' also putting heavy emphasis on the need for drivers to remain engaged with hands on the wheel while using the ``Autopilot'' feature~\citep{the_tesla_team_tragic_2016}.

While it acknowledged the role of human errors (both the truck driver and the Tesla driver), the National Transportation Safety Board (NTSB) determined that ``the operational design of the Tesla’s vehicle automation permitted the car driver’s overreliance on the automation, noting its design allowed prolonged disengagement from the driving task and enabled the driver to use it in ways inconsistent with manufacturer guidance and warnings.''~\citep[p.42]{the_national_transportation_safety_board_collision_2017}
In addition, it found that ``[t]he Tesla’s automated vehicle control system was not designed to, and could not, identify the truck crossing the Tesla’s path or recognize the impending crash.'' (p.30)

While this verdict does not directly specify what safety constraints were missing or unsuccessfully enforced, it does provide suggestions.
Clearly, if the autonomous driving feature continues, there need to be put in place better conditions to prevent prolonged disengagement and unsafe use.
The faulty vision system needs a structural reconceptualization to either ensure that predictable scenarios like a truck crossing are detectable or to constrain the use of the vision system only to situations where its errors cannot lead to hazardous conditions. 



\subsection{Shift from a Probabilistic to a System-theoretic Safety Perspective}
\label{sec:mechanisms}
\noindent \textit{\textbf{Leveson Lesson 3:} 
Risk and safety may be best understood and communicated in ways other than probabilistic risk analysis.}

Most AI systems parameters are optimized according to some objective or reward function.
These tend to either minimize or maximize a probability that the AI model output matches a desirable outcome. 
It is no secret that such approaches will never \emph{guarantee} that an AI system's output enters unsafe territory- they may only reduce the probability.
Just as probabilistic risk assessments, the reward or objective function tends to absorb a plethora of risky dynamic safety scenarios into one function without explicitly formulating these scenarios, let alone how to safeguard against these.

Internalizing constraints in the design and training of ML-driven AI systems is notoriously hard and, while some relevant efforts have been made in theory~\citep{achiam_constrained_2017}, a workable and scalable solutions is not yet available.
Solely relying on an AI system's internal model to incorporate all safety constraints is dangerous. \citet{raji_concrete_2020} note that you can expect AI system models to fail and hence, of true scrutiny should be the effectiveness of the \emph{safety measures} employed, such as fail-safe features to kick in when an anticipated engineering failure occurs.
Instead of capturing safety in terms of low probabilities, alternative methods are needed to secure a system in practice, by combining learning-methods with control-theoretic guarantees~\citep{fisac_general_2019} and installing other (socio-technical) fail-safe mechanisms~\citep{dobbe_hard_2021}. 
To design those effectively, we first have to reevaluate where we draw the boundary of our system design and accompanying safety analysis.

\subsubsection*{Relevant System Safety Strategy: Capture the safety conditions and assumptions in a process model}

\begin{figure}[!htb]
    \center{\includegraphics[width=14cm]
    {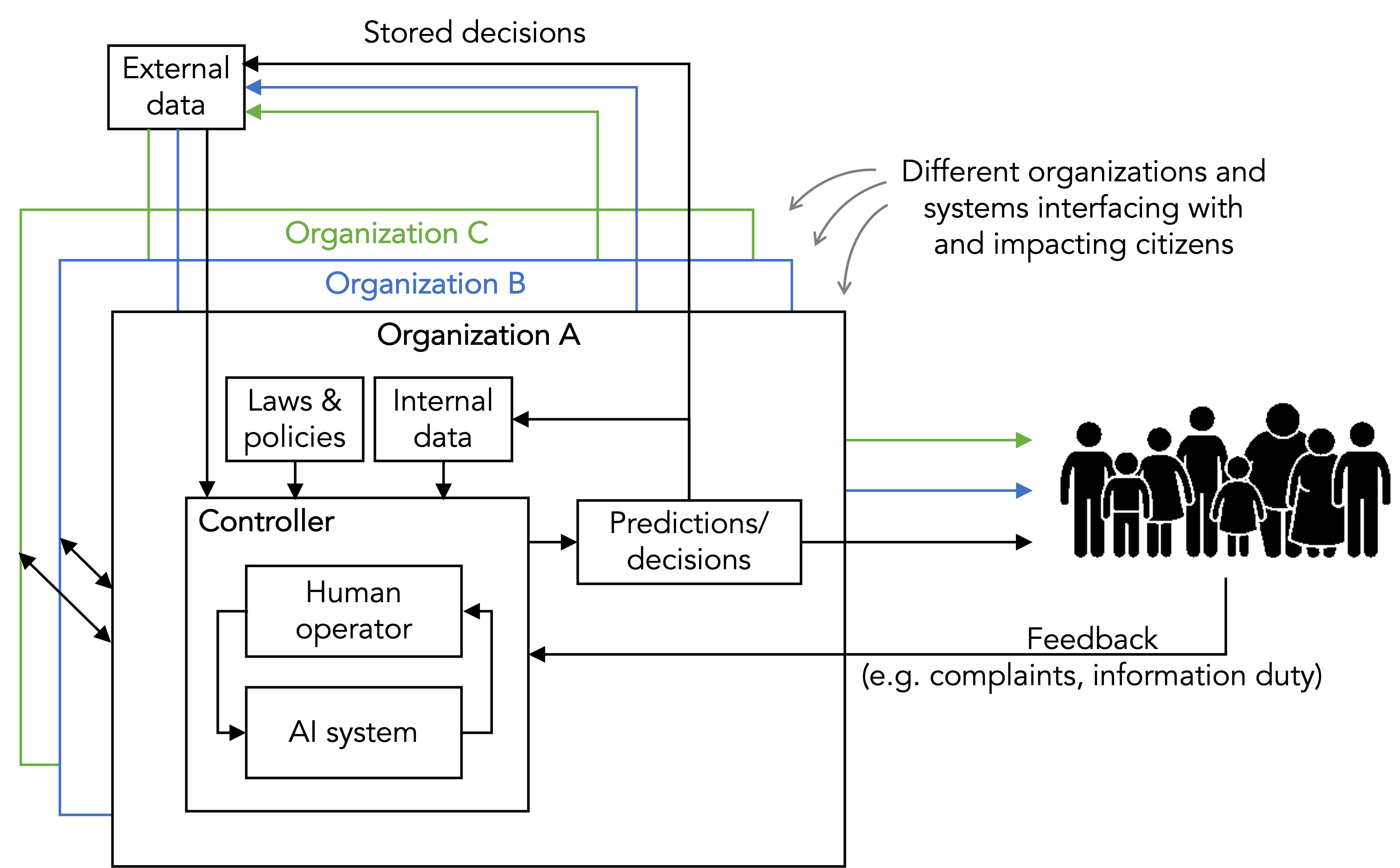}}
    \caption{\label{fig:processmodel} A specific depiction of the process model in the context of several government organizations making AI-informed decisions about households that are dynamically linked through various data registers.}
\end{figure}

The boundary of analysis for safety should extend beyond the AI model (with its inputs, features, outputs and reward/objective function), and include dynamics of the AI model interacting with other system components, including human operators, users or subjects and the process and environment the system is intervening in.
In system safety, this is done with a \emph{process model}, which interprets an AI system as a control entity, possibly complemented by forms of human control and decision-making, as indicated in Figure~\ref{fig:processmodel}. 
The process model formulates four conditions:
\begin{enumerate}
    \item \textit{The goal:} the objectives and safety constraints that must be met and enforced by the controller;
    \item \textit{The action condition:} the controller must be able to affect the state of the system;
    \item \textit{The observability condition:} the controller must be able to ascertain the state of the system, through feedback, observations and measurements;
    \item \textit{The model condition:} the controller must be or contain a model of the process. A human controller should also have a model of the behavior of the AI techniques used for control and decision-making.
\end{enumerate}

Based on the process model, it is easier to specify what feedback and fail-safe mechanisms are needed to enforce safety constraints, and, as a consequence, to understand how accidents occur. 

Leveson argues that often, accidents occur ``when the process model used by the controller (automated or human) does not match the process.''~\citep[p.88]{leveson_engineering_2012}
This can lead to four types of control errors: incorrect or unsafe control actions, missing control actions for guaranteeing safety, wrong timing of control actions, or control actions applied too long or stopped too soon. 

Here we end with an example from the public sector in The Netherlands.
The Dutch government developed the System Risk Indication (SyRI); a legal instrument to detect various forms of fraud, including social benefits, allowances, and taxes fraud, by developing risk profiles using statistical learning techniques.
These risk profiles determined the deployment of resources to combat fraud.
In 2020, The Hague District Court ruled the SyRI system and its accompanying legislation unlawful as it violated the European Convention on Human Rights (ECHR): ``The court finds that the SyRI legislation in no way provides information on the factual data that can demonstrate the presence of a certain circumstance, in other words which objective factual data can justifiably lead to the conclusion that there is an increased risk.''~\citep[article 6.87]{rechtbank_den_haag_eclinlrbdha20201878_2020}
Later that year, the Dutch Data Protection Agency concluded that risk classification system had been used in the childcare benefits scandal, in which more than twenty-six thousand families were wrongfully accused of fraud~\citep{autoriteit_persoonsgegevens_verwerking_2020}.
Many families collected steep amounts of debt and were forced to foreclose and sell their homes or vehicles, also leading to massive emotional harms including forced divorces and even cases of suicide~\citep{levie_dutch_2021}.

While the scandal is still under investigation, it is clear that AI systems played a central role in causing these harms, with algorithms discriminating based on ethnicity and income level.
This case exemplifies the sheer lack of a process model: The system informed decisions (action condition) based on illegitimate inputs (observation condition) and had no idea of the detrimental consequences for citizens (model condition) nor did it have any safeguards in place (goal condition).
The AI system used was part of a broader policy and governance structure that dehumanized thousands of families.
Arguably, the model of the controlled process should have included the impact of possible control errors (in this case the incorrect classification of families as `fraudulent') on the financial and general health of the subject families.
That would have enabled the formulation of safety constraints that would rule out the unjustified condemnation.
In the Dutch legal system, this would have included proper avenues for legal protection and due process in the case of disagreement on the verdict of the system and its operators.
Such safety hazards would have been naturally identified if human dimensions were part of the model of the controlled process, which is now a top priority for many policy domains.
%
The Dutch case shows that their public administration, while being leading in terms of digitalization, does not have a long history of dealing with safety hazards and centering (possible) victims of control errors in improving automated decision-making.
As such, it will be able to learn significantly from the lessons learned in system safety.

\subsection{Shift from Siloed Design and Operation to Aligning Mental Models}
\label{sec:operator}
\noindent \textit{\textbf{Leveson Lesson 4:} Operator error is a product of the environment in which it occurs. To reduce operator ``error'' we must change the environment in which the operator works.
}

While investigations into accidents tend to structurally find other factors, there is a recurring tendency is to put the blame on inadequate response to the failure by an operator~\citep{leveson_engineering_2012}.
This phenomenon has been attributed to the effect of \emph{hindsight bias} in which the focus is on what an operator did wrong rather than on what was the logical thing to do given the circumstances~\citep{dekker_just_2016}. 
In the context of AI systems research, the dominant focus on technical solutions often contribute to an under-appreciation for the role of human operators and other users of a system~\citep{dobbe_hard_2021}.

Meanwhile, new regulatory proposals are putting more emphasis on the need for human oversight, as a response to emerging evidence on the dangers of AI systems. 
The \citet{european_commission_proposal_2021} highlights the role of meaningful human oversight in their recent proposal for regulating AI systems.
\citet{kak_false_2021} point out that this may not address societal concerns, showing how earlier calls have provided shallow protection in the past and may instead serve as a means of ``rubber stamping'' intrusive and harmful applications, thereby blurring responsibility, where frontline human operators of A.I. systems are blamed for broader system failures over which they have little or no control
~\citep{elish_moral_2019}.


\subsubsection*{Relevant System Safety Strategy: Align mental models across design, operation and affected stakeholders}

While issues around the operator and human-machine interactions occurring with new AI systems seem new, these have been studied extensively for safety-critical systems such as airplane autopilots.
Here I focus on the importance of \emph{mental models}, which are maintained by designers and operators and ``used to determine what control actions are needed, and it is updated through various forms of feedback''~\citep[p.87]{leveson_engineering_2012}. 
Mental models are a concept that finds its roots in cognitive science and engineering and the study of work performance~\citep{rasmussen_mental_1987}.

Across the mental models formed by designers and operators natural changes emerge over time.
Operators or users typically find rational grounds to deviate from the \emph{normative procedures} described by designers, leading to \emph{effective procedures}~\citep{rasmussen_cognitive_1994}.
This \emph{adaptation} is common across all systems and may happen due to various reasons, such as time constraints, economic or political incentives, or an operator understanding how to better safeguard or optimize a system through first-hand experience, which a designer typically lacks.
It is therefore no surprise that operator error is found as a dominant focus for accident causes, as any deviation from normative procedure could be labeled as erroneous.
In turn, this effect is often used as an invalid argument to automate the operator away~\citep{dekker_field_2017}.

\begin{figure}[!htb]
    \center{\includegraphics[width=10cm]
    {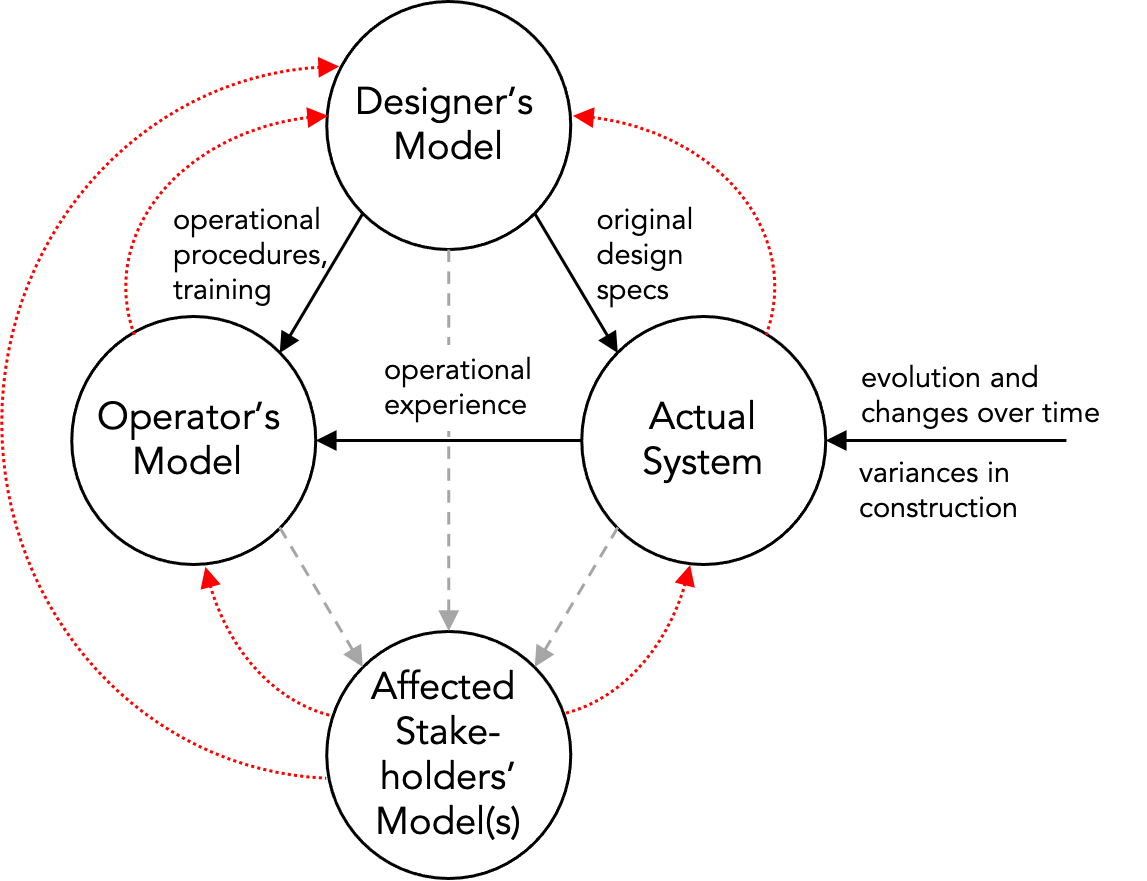}}
    \caption{\label{fig:mentalmodels} Relationships between mental models, adapted from~\citep{leveson_engineering_2012}. }
\end{figure}

Instead, the system safety discipline would describe efforts to align and periodically update the different \emph{mental models} that different actors hold and use when designing, operating or otherwise interacting with the system. 
In Figure~\ref{fig:mentalmodels}, I show an adaptation of Leveson's depiction of the relationships between mental models (straight arrows). 
In addition to designers and operators, the increasing dominance of automation in societal infrastructures also leads to other stakeholders forming mental models of a system's behavior and how to depend on and interact with it. 
The updating of mental models through feedback mechanisms will be further addressed in Section~\ref{sec:operations}.

Here we discuss the importance of mental models for a recent case in autonomous driving.
The fatal accident in which an autonomous vehicle, developed and tested by Uber, killed a road-crossing pedestrian in Tempe, Arizona, on March 18, 2018, forms a painful example. 
When evidence emerged that the test driver, responsible for taking over control of the car in the event of unforeseen and risky conditions, had been visually distracted due to looking at her cellphone, most public journalism focused on her role in causing the accident.
But while the NTSB did identify the operator's role as the probable cause of the accident, it also put major emphasis on the role Uber had played in enabling such unsafe behavior: ``Contributing to the crash were the Uber Advanced Technologies Group’s (1) inadequate safety risk assessment procedures, (2) ineffective oversight of vehicle
operators, and (3) lack of adequate mechanisms for addressing operators’ automation complacency—all a consequence of its inadequate safety culture.''~\citep[p. v]{the_national_transportation_safety_board_collision_2019} 

The damning NTSB verdict makes clear the essence of aligning the mental models across design and operation.
While it is evident that the behavior the test driver exhibited was unsafe, there were various environmental factors Uber could have addressed to prevent this from happening. 
A typical reflex to this is to install human factor checklists for designers and put in place stricter procedures for operators to follow. These however, have been shown not to be sufficient in establishing high levels of safety, and can even have the opposite effect~\citep{dekker_field_2017}.
Instead, engineers and designers need to design the system and the environment to prevent human errors.
A general approach to designing adequate shared control of humans and AI systems is unrealistic as there are a plethora of contextual considerations~\citep{leveson_engineering_2012}. 
Nevertheless, Leveson has outlined three principles:
\begin{enumerate}
    \item \textit{Design for redundant paths:} to provide multiple paths to ensure that a single error cannot prevent the operator from taking action to maintain a safe system state and avoid hazards;
    \item \textit{Design for incremental control:} to give the operator enough time and feedback to perform control actions and, if possible, to do so in incremental steps rather than in one control action;
    \item \textit{Design for error tolerance:} to make sure that reversible errors are observable to the human operator before unacceptable consequences occur, to allow them to monitor their own performance and to recover from erroneous actions.
\end{enumerate}


\subsection{Curb the Curse of Flexibility in AI Software Development}
\label{sec:software}
\noindent \textit{\textbf{Leveson Lesson 5:} Highly reliable software is not necessarily safe. Increasing software reliability or reducing implementation errors will have little impact on safety.
}

Perhaps the most pertinent and under-appreciated lesson from Leveson's system safety canon relates to the role of software in algorithmic harms. The \emph{curse of flexibility} refers to the dominant assumption that, in contrast to physical machines, software is not subject to physical constraints, leading to ever more flexible and complex designs aiming to address existing and emerging requirements.
However, Leveson argues that the limiting factors of software-based automation change from the structural integrity of materials to limits on people's intellectual capabilities~\citep{leveson_safeware_1995,leveson_engineering_2012}.


The source of most serious problems with software relate to outsourcing software development. 
This creates extra communication steps between those understanding the physical realization of the system (domain experts, users, affected stakeholders, hardware operators) and those programming the software. 
And this step often leads to requirements not being complete. 
Leveson argues that ``[n]early all the serious accidents in which software has been involved in the past twenty years can be traced to requirements flaws, not coding errors. [..] The most serious problems arise, however, when nobody understands what the software should do or even what it should not do.''~\citep[p.49]{leveson_engineering_2012}
Unfortunately, there is mounting evidence that the development of high-stakes AI systems and software often goes without principled ways to determine safety requirements and translate these to software implementations~\citep{dobbe_hard_2021}.

The resulting lack of proper requirements and validation is best reflected in the development of large language models (LLMs) within the field of natural language processing (NLP). 
LLMs are notorious for their complexity and flexibility.
This combined with the struggle to impose reasonable safety requirements make these systems unsafe by design, with large environmental consequences~\citep{dobbe_ai_2019}, damage to cultural heritage and harm to groups and individuals subject to an LLM's errors~\citep{bender_dangers_2021}.

\subsubsection*{Relevant System Safety Strategy: Include software and related organizational and infrastructural dependencies in system-theoretic hazard analysis}

The System-Theoretic Process Analysis (STPA) that Leveson developed formalized hazard analysis to provide information and documentation necessary to ensure the enforcement of safety constraints in system design, development, manufacturing, and operations~\citep{leveson_engineering_2012}.
The output of such analysis can be used to develop or update the requirements of a software-based automation system in the process model (as presented in Section~\ref{sec:mechanisms}), as well as the institutional arrangements in the safety control structure.

STPA considers software as a primary source for possible hazard, as it modulates decisions and control actions, provides feedback (e.g. through sensor readings) and executes computational models. 
A more nascent application of STPA is to use it to identify safety risks related to how the software is organized.
Over the last decade, we have seen a major shift to \emph{cloud computing platforms}, in which a broad variety of services for software development and maintenance are integrated, including tools for integrating data and building AI applications.
Concurrently, AI applications have been championed to solve complex problems in marketing efforts by vendors, both those developing cloud platforms and third parties working with it. 
The AI Now Institute showed how the AI hype has contributed to ``widening gaps between marketing promises and actual product performance. With these gaps come increasing risks to both individuals and commercial customers, often with grave consequences.'' ~\citep[p.5]{whittaker_ai_2018}

\citet{gurses_privacy_2017} show how cloud computing platforms and their central bundling of services have challenged principled software development, rendering designing for crucial values like safety or privacy in AI system software an elusive goal. 
More alarmingly, the control of big tech companies over infrastructures central to development of high-risk and high-impact AI systems threatens our ability to understand and regulate these technologies~\citep{whittaker_steep_2021}.

As a result, there is a set of open problems related to safeguarding the organizational and infrastructural risks of software used to build AI systems in critical public domains, in ways that overcome and rebalance power relations~\citep{balayn_beyond_2021}. Here, a system safety can help us to expand the boundary of safety analyses to include organizational and infrastructural dependencies; to directly address the hazards that emerge through contracts that lean on central bundling of software services; and to characterize the risks that emerge from not being able to program safety requirements in an end-to-end fashion.

\subsection{Translate Safety Constraints to the Design and Operation of the System}
\label{sec:operations}
\noindent \textit{\textbf{Leveson Lesson 6:} Systems will tend to migrate toward states of higher risk. Such migration is predictable and can be prevented by appropriate system design or detected during operations using leading indicators of increasing risk.
}


Through a safety-guided design, many hazards may be anticipated and eliminated. 
However, there are various reasons why hazards may arise during operations.
After decades of research, system safety pioneer Jens Rasmussen concluded that systems tend to migrate toward states of higher risk, and that such adaptation is not random but predictable and hence manageable~\citep{rasmussen_risk_1997}.
An important recurring observation in his research shows that safety defenses are likely to degenerate, especially ``when pressure toward cost-effectiveness is
dominating,'' and that often accident investigations conclude that the particular accident was ``waiting for its release''. (p.189)
The \emph{safety control structure} should hence assign responsibilities for monitoring and overseeing that the system operates safely subject to inherent adaptation.

\subsubsection*{Relevant System Safety Strategy: Organize feedback mechanisms for operational safety}

An operational safety control structure establishes controls and feedback loops to ``(1) identify and handle flaws in the original hazard analysis and system design and (2) to detect unsafe changes in the system during operations before the changes lead to losses.''~\citep[p.394]{leveson_engineering_2012}
Here I focus on the value of three concrete feedback mechanisms for maintaining and improving operational safety, namely audits, accident investigations and reporting systems.
Note that these mechanisms are different from the feedback that the controller (a combination of the AI system and/or a human operator) receives in real-time to control a process (as depicted in Figure~\ref{fig:processmodel}). 
Instead, this feedback forms input to reevaluate the system design, to update safety constraints (based on hazard analysis) and to update the process model and roles and responsibilities in the safety control structure.

Audits and performance assessments are done to determine whether safety constraints are enforced in the operation of the system, and whether the rationale of the system design holds up in practice.
The need for auditing AI systems is still in its infancy, but there have been some efforts to audit for the presence of bias~\citep{raji_actionable_2019}.
Effectuating functional audits may be a struggle, as behaviors may change in anticipation of an audit or a lack of independence may influence its objectivity.
Overcoming such issues requires building a conducive organizational culture (see Section~\ref{sec:organization}).
A \emph{participatory audit} aims to impact the cultural challenges~\citep[Chapters 12 and 14]{leveson_engineering_2012}, partly by not making audits punitive and ensuring all levels of the safety control structure are audited, including affected stakeholders.
This aspect is corroborated by Rasmussen's findings, who stresses the need ``to represent the control structure involving all levels of society for each particular hazard category''~\citep[p.183]{rasmussen_risk_1997}.

Accident, incident and anomaly investigations are meant to draw lessons and potential conclusions for improving the system design, process model and safety control structure.
These investigations have historically suffered from \emph{root cause seduction}, meaning a too narrow focus on finding root causes to explain complex accidents~\citep{carroll_incident_1995}.
Instead, as discussed in Section~\ref{sec:accidentmodel}, investigations should use a systems perspective for understanding accidents.
A crucial component is to develop such a perspective in the organizations so that investigations can be done holistically and lead to follow-up and training of all involved professionals.

Finally, reporting systems can help to detect issues before they turn into hazards.
Reporting systems are common in safety-critical domains, such as healthcare or aviation.
As discussed for the other feedback mechanisms, there are hurdles to ensuring a reporting system is used. 
It is therefore important to understand the factors why people won't use them and fix those.
Common factors have to do with an ineffective or unclear interface of the system, a lack of follow-up, or fear that reported information may be used against the person or someone else.


\subsection{Build an Organization and Culture that is Open to Understanding and Learning}
\label{sec:organization}


\noindent \textit{\textbf{Leveson Lesson 7:} Blame is the enemy of safety. Focus should be on understanding how the system behavior as a whole contributed to the loss and not on who or what to blame for it.
}

As described earlier, many major accidents in complex systems are met with subjectivity.
Safety is a deeply normative concept that is understood, formalized and experienced in different ways across different cultures, communities and organizations.
This subjectivity is a major source of challenges in the emerging research and practices for safeguarding AI systems~\citep{dobbe_hard_2021}.
For example, the attribution of accident causes can be subject to the characteristics of the victim and the analyst, such as hierarchical level, job satisfaction and level involvement in the accident~\citep{leplat_occupational_1984}.

The emerging harms related with AI systems have motivated a plethora of ethical principles, guidelines and policy instruments~\citep{schiff_ai_2021}.
While safety is a core tenet of recent regulatory efforts~\citep{european_commission_proposal_2021}, the accountability gaps existing for many AI-driven innovations~\citep{whittaker_ai_2018} may be a source of blame games and finger pointing that can take away attention from understanding how accidents may happen or have happened as a way to do better at building and operating AI systems.
While sometimes misinterpreted as a means to divert accountability from humans to the system, the establishment of system safety has contributed greatly to new more effective ways to promote and structure individual responsibility and accountability~\citep{dekker_systems_2015}, especially when combined with cultural change programs that promote sharing of information to improve safety, as evidenced in the healthcare
and
aviation industry~\citep{dekker_just_2016}.

\subsubsection*{Relevant System Safety Strategy: Balancing safety and accountability through a Just Culture}

Leveson's last lesson is the most crucial. 
In the final chapters of her book \emph{Engineering a Safer World: Systems Thinking Applied to Safety}, Leveson underlines the importance of adequate management and safety culture to accomplish any of the goals described in the previous sections\citep{leveson_engineering_2012}.
The increasing scale and complexity of AI systems, makes that these challenges span broader institutional networks, often comprising public, private, knowledge and societal institutions~\citep{janssen_challenges_2016}.

Leveson outlines the core organizational requirements managers need to meet for improving safety; an effective safety control structure (as discussed in Sections~\ref{sec:systemhazards} and~\ref{sec:operations}), a safety information system, and a strong and sustainable safety culture. Here we focus on the often under-appreciated importance of the latter.

Bundled in his book \textit{Just Culture: Balancing Safety and Accountability}, Sidney Dekker describes how to stray away from detrimental blame cultures, instead arguing for the crucial need to provide a comfortable environment in which people feel safe to share information about what should be improved to those who can do something about it, and to allow for investments in improvements and learning that have a real safety dividend rather than spending money on legal protection and limiting liability~\citep{dekker_just_2016}.
Through his extensive field work and experience as a pilot, Dekker explains the struggles that organizations go through to build an effective safety culture. 
Unfortunately, there is no simple formula, but Dekker has gained important insights into \emph{how} to establish such cultures and core agreements on system safety procedures and responsibilities: ``What matters for organizational justice is not so much whose version gets to rule the roost— but to what extent this is made explicit and how it is decided upon.''~\citep[p.75]{dekker_just_2016} The key steps he outlines are:
\begin{enumerate}
    \item Design the process to deal with adverse events or apparently risky acts, which does not interfere with performance review and makes explicit how judgments are made and enables opportunities for appeal;
    \item Decide who is involved in this process, in a way that allows for impartial input from across the organization and includes domain expertise to represent and acknowledge work floor complexity;
    \item Decide who is involved in deciding who is involved, to prevent top-down decision-making that lacks buy-in and ownership.
\end{enumerate}

The aviation industry is historically known for its safety culture, as also studied by Dekker~\citep{dekker_just_2016}.
In recent years, two crashes occurred with a Boeing 737-MAX, in which the plane took an uncontrollable nosedive.
It turned out that something had gone wrong with the aircraft calibration algorithm~\citep{gelles_i_2020}.
Driven by competitive forces to update the 737 models with new engine technology, the company decided to bypass certain checks on how new engines would affect the autopilot behavior, ultimately causing pilots to be confused and unable to steer the plane out of a nosedive.
As such, despite the broader industry's emphasis on safety culture, Boeing's own culture had been a breeding ground for irresponsible decisions and unsafe design choices~\citep{boeing_boeing_2019}.
This example shows the detrimental effects that organizational culture and decision-making can have on engineering choice, and in effect the safety of a system in operation.
Hence, organizations responsible for such systems should build a strong safety culture, and have the leadership and mechanisms in place to maintain it.


\begin{table}
    \centering
    \begin{tabular}{c | p{0.2\linewidth} | p{0.325\linewidth} | p{0.325\linewidth}}
        &\multicolumn{1}{c|}{\textbf{Leveson Lesson}} & \multicolumn{1}{|c|}{\textbf{AI System Safety Implication}} & \multicolumn{1}{|c}{\textbf{Example System Safety Strategy}}  \\
         \hline 
         1 & Component reliability is insufficient for safety & Identify and eliminate hazards at system level & System hazard-informed system design and safety control structure \\ \hdashline
         2 & Causal event models cannot capture system complexity & Understand safety through socio-technical constraints & System-theoretic accident models: integrating safety constraints, the process model and the safety control structure \\ \hdashline
         3 & Probabilistic methods don't provide safety guarantees & Capture safety conditions and requirements in a system-theoretic way  & Process model: AI system goals, actions, observation and model of controlled process and automation  \\ \hdashline
         4 & Operator error is a product of the environment & Align mental models across design, operation and affected stakeholders & Leveson's design principles for shared human-AI controller design: redundancy, incremental control and error tolerance \\ \hdashline
         5 & Reliable software is not necessarily safe & Include (AI) software and its organizational dependencies in hazard analysis & System-theoretic process analysis \\ \hdashline
         6 & Systems migrate to states of higher risk & Ensure operational safety & Feedback mechanisms (audits, investigations and reporting systems) \\ \hdashline
         7 & Blame is the enemy of safety & Build an  organization and culture that is open to understanding and learning & Just Culture 
    \end{tabular}
    \label{tab:overview}
    \caption{Overview of Leveson lessons, implications for AI system development and governance and examples of relevant system safety strategies.}
    \label{tab:my_label}
\end{table}

\section{Conclusions}


In this chapter, I covered central insights from the field of system safety, with particular focus on what it takes to safeguard systems that are subject to software-based automation.
The lessons and concrete system safety strategies are summarized in Table~\ref{tab:overview}.
We have seen that most of these lessons have yet to be absorbed in domains where AI systems are emerging to inform, mediate or automate decision-making.

It would however be naive to conclude that the strategies presented here can be straightforwardly applied.
AI systems' inherent flexibility and associated failure modes make designing for safety, as well as formulating effective process models and safety control structures a complex task. 
If anything, the power of evidence that system safety strategies have could help understand when using an AI system is a good idea to begin with. 
Formulating constraints for valid safety problems can bring sanity to what is ``responsible AI'' by making explicit when and how not to use it, in order to prevent unnecessary harm. 

Lastly, it is crucial to note that any tool or strategy can be applied in irresponsible ways, system safety approaches included.
It should be evident from this chapter, that applying these strategies requires various forms of formalization and modeling.
As such, system safety methods will always redraw, solidify and impose power relationships.
This is not necessarily a bad thing, but should be made explicit in an effort to empower those that need safeguarding.
Just as patient safety has become the central focus in successful healthcare systems, vulnerable people and communities should have a seat at the table and see themselves empowered in the crystallization of a system's design and operational safety control structure.

\bibliography{references_oxford}
\bibliographystyle{apalike}

\end{document}